\documentclass[10pt]{article}
\usepackage{cite}
\usepackage{epsfig}
\usepackage[dvips,usenames]{color}
\usepackage{color}
\usepackage{rotating}
\usepackage{amsmath,amssymb,amsfonts}
\usepackage{latexsym}
\begin{document}
\setcounter{section}{0}
\setcounter{equation}{0}
\setcounter{figure}{0}
\setcounter{table}{0}
\setcounter{footnote}{0}
\begin{center}
{\bf\Large The $1+1$ Dimensional Abelian Higgs Model Revisited:}

\vspace{5pt}

{\bf\Large Non-perturbative Dynamics in the Physical Sector}\footnote{Contribution to the Proceedings of the Fifth International
Workshop on Contemporary Problems in Mathematical Physics, Cotonou, Republic of Benin, October 27--November 2, 2007,
eds. Jan Govaerts and M. Norbert Hounkonnou (International Chair in Mathematical Physics and Applications,
ICMPA-UNESCO, Cotonou, Republic of Benin, 2008), pp.~159--163.}
\end{center}
\vspace{10pt}
\begin{center}
Laure GOUBA\\
\vspace{5pt}
{\sl National Institute for Theoretical Physics (NITheP),\\
Stellenbosch Institute for Advanced Study (STIAS),\\
Private Bag X1, Matieland 7602, Republic of South Africa}\\
{\it E-Mail: gouba@sun.ac.za}\\
\vspace{7pt}
{\sl African Institute for Mathematical Sciences (AIMS),\\
6 Melrose Road, 7945 Muizenberg, Republic of South Africa}\\
{\it E-Mail: laure@aims.ac.za}\\
\vspace{10pt}
Jan GOVAERTS\\
\vspace{5pt}
{\sl Center for Particle Physics and Phenomenology (CP3),\\
Institut de Physique Nucl\'eaire, Universit\'e catholique de Louvain (U.C.L.),\\
2, Chemin du Cyclotron, B-1348 Louvain-la-Neuve, Belgium}\\
{\it E-Mail: Jan.Govaerts@uclouvain.be}\\
\vspace{7pt}
$^\star${\sl Fellow, Stellenbosch Institute for Advanced Study (STIAS),\\
7600 Stellenbosch, Republic of South Africa}\\
\vspace{7pt}
{\sl International Chair in Mathematical Physics and Applications (ICMPA-UNESCO Chair),\\
University of Abomey--Calavi, 072 B. P. 50, Cotonou, Republic of Benin}
\end{center}

\vspace{15pt}

\begin{quote}
In this paper the two dimensional Abelian Higgs model
is revisited. We show that in the physical sector, this model
describes the coupling of the electric field to the radial part, in field
space, of the complex scalar field.
\end{quote}

\vspace{10pt}

\section{Introduction}
\label{Sec1}

Two dimensional scalar electrodynamics is the coupling
of a U(1) gauge field to a scalar complex field, where the potential
can be chosen arbitrarily. Associated to the Higgs potential, the model
corresponds to the well-known abelian Higgs model which has been used
to illustrate spontaneous gauge symmetry breaking, or provides an effective
description for superconductivity through the Landau--Ginzburg model in
four dimensional Minkowski spacetime. In this contribution we show \cite{lau}
that in its physical gauge invariant sector this model corresponds
to the coupling of the electric field to the radial
part---in field space---of the initial complex scalar field.

This paper is organised as follows. In Section~\ref{Sec2}, the two dimensional
abelian Higgs model is introduced. In Section~\ref{Sec3}, the non-perturbative dynamics in the 
physical sector of the system is identified. Concluding remarks
are provided in Section~\ref{Sec4}.

\section{The Two Dimensional Abelian Higgs Model}
\label{Sec2}

\subsection{Choice of Lagrangian}

Let us consider a two dimensional Minkowski spacetime, where the metric $\eta_{\mu\nu}$ ($\mu,\nu=0,1$) is of
signature $\eta_{\mu\nu}={\rm diag}\,(+-)$. The coupling of a gauge field $A_\mu$ to the complex
scalar field $\phi$ is characterised by the gauge coupling constant $e$, while the scalar field $\phi$
possesses self-interactions associated to a U(1) invariant potential $V(|\phi|)$. By construction this U(1)
symmetry defines the associated gauge group. The Lagrangian density of this model is,
\begin{eqnarray}
\nonumber
{\mathcal{L}} =-\frac{1}{4}\left[\partial_\mu A_\nu -\partial_\nu A_\mu \right]
\left[\partial^\mu A^\nu -\partial^\nu A^\mu\right] +
\vert\left(\partial_\mu + ie A_\mu\right)\phi \vert^2 - V(|\phi|),
\end{eqnarray}
where it is assumed without loss of generality that $e>0$. More explicitly, by expanding the implicit summations
over the repeated spacetime indices, one has
\begin{equation}
\label{401}
{\mathcal{L}} = \frac{1}{2}\left(\partial_0 A_1 -\partial_1 A_0\right)^2
+\left[\partial_0\phi^{*} -ieA_0\phi^{*}\right]
\left[\partial_0\phi + ieA_0\phi\right]
- \left[\partial_1\phi^{*} -ie A_1\phi^{*}\right]
\left[\partial_1\phi + ie A_1\phi \right]
-V(|\phi|).
\end{equation}

In order to conveniently factorise the gauge invariant and variant
contributions to the dynamics, let us introduce a polar parametrisation of 
the complex scalar field as follows,
\begin{eqnarray}
\label{400}
\phi(x^\mu)=\frac{1}{\sqrt 2}\rho(x^\mu)e^{i\varphi(x^\mu)},
\end{eqnarray}
where the factor $1/\sqrt{2}$ is a convenient choice of 
normalisation. It is understood that the functions $\rho(x^\mu)$ 
and $\varphi(x^\mu)$ are continuous on the two dimensional Minkowski spacetime, 
the sign of $\rho(x^\mu)$ being arbitrary. However the choice for the sign of $\rho(x^\mu)$
is correlated to the choice of the evaluation for $\varphi(x^\mu)$
modulo $\pi$ as
\begin{eqnarray}
\nonumber
\left[(-1)^k\rho, \varphi +k\pi \right]
\longleftrightarrow \left[\rho,\varphi\right],\qquad k\in \mathbb{Z}.
\end{eqnarray}
Furthermore,  $\varphi(x^\mu)$ can be multi-valued at the zeros of $\rho(x^\mu)$.
We thus also need to take into account for the gauge invariance of physical configurations their
invariance properties under the associated $\mathbb{Z}_2$ symmetry,
$\rho \leftrightarrow -\rho$.

Substituting (\ref{400}) into (\ref{401}), we have
\begin{equation}
{\mathcal{L}} = \frac{1}{2}\left[\partial_0 A_1 - \partial_1 A_0\right]^2 
+\frac{1}{2}\left(\partial_0\rho\right)^2 +
\frac{1}{2}\rho^2\left(\partial_0\varphi +e A_0\right)^2
- \frac{1}{2}\left(\partial_1\rho\right)^2 -\frac{1}{2}\rho^2
\left(\partial_1\varphi + e A_1\right)^2 - V(\rho).
\end{equation}

\subsection{Hamiltonian formulation}

The degrees of freedom of the model in its Lagrangian description are
\begin{eqnarray}
A_0(t,x),\:A_1(t,x),\:\rho(t,x),\:\varphi(t,x). 
\end{eqnarray}
Note that the electric field is given by
\begin{eqnarray}
E =-\partial_1A^0 -\partial_0A^1 =\partial_0A_1 -\partial_1A_0 =F_{01}.
\end{eqnarray}
The conjugate momenta associated to these degrees of freedom are,
respectively,
\begin{eqnarray}
\pi^0 = 0,\qquad
\pi^1 = E,\qquad
\pi_\rho = \partial_0 \rho,\qquad
\pi_\varphi = \rho^2\left(\partial_0\varphi + e A_0\right).
\end{eqnarray}
The associated non-vanishing Poisson brackets are given by,
\begin{eqnarray}
\nonumber
\left\{ A_0(t,x),\,\pi^0(t,y)\right\} &=&\delta (x-y) = -\left\{\pi^0(t,x),\, A_0(t,y)\right\}, \nonumber \\
\left\{ A_1(t,x),\,\pi^1(t,y)\right\} &=&\delta (x-y) = -\left\{\pi^1(t,x),\, A_1(t,y)\right\},\nonumber \\
\left\{ \varphi(t,x),\,\pi_\varphi(t,y)\right\} &=&\delta (x-y) = -\left\{\pi_\varphi(t,x),\, \varphi(t,y)\right\}, \\
\left\{ \rho(t,x),\,\pi_\rho(t,y)\right\} &=&\delta (x-y) = -\left\{\pi_\rho(t,x),\, \rho(t,y)\right\}. \nonumber
\end{eqnarray} 
Given these degrees of freedom, their conjugate momenta and the associated
Poisson brackets, the dynamics of the system also derives from the following canonical Hamiltonian
\begin{eqnarray}
\nonumber
H_0 &=& \int dx \left(
\frac{1}{2}\left(\pi^1\right)^2 +
\frac{1}{2}\left(\pi_\rho\right)^2 +
\frac{1}{2}\left(\partial_1\rho\right)^2 +
\frac{1}{2}\rho^2\left(\partial_1\varphi + e A_1\right)^2 \right.\\\label{408}
 &+&\left.
\frac{1}{2}\frac{1}{\rho^2}\pi_\varphi^2   
 + \partial_1 (A_0\pi^1)
  - A_0\left(\partial_1\pi^1 + e\pi_\varphi\right)   
+ V(\rho)
\right),
\end{eqnarray}
where the range of integration for the spatial variable is to be specified shortly.

\section{Non-perturbative dynamics in the physical sector}
\label{Sec3}

Let us now choose to compactify the real spatial line into a circle of length $2L$, with $-L\le x\le L$.
Hence the topology of spacetime becomes that of a cylinder, $\tau = \mathbb{R}\times S^1$.
Such a choice improves the Fourier mode analysis of the degrees of freedom of the system
and their subsequent quantisation. Moreover, being gauge invariant the model possesses
a primary constraint. Indeed the momentum $\pi^0$ conjugate to $A_0$ leads to the primary constraint
$\sigma_0  =\pi^0=0$. Hence one needs to study the dynamics of this constraint to see whether
other constraint are generated. Indeed there should exist at least another one, namely Gauss' law constraint
for this electromagnetic interaction. Using the Dirac formalism for constraints \cite{Gov1},
it is readily shown that the primary constraint $\sigma_0$ generates only one more
secondary constraint, $\sigma_1$, namely Gauss' law; that the latter constraint does not generate any
further constraint; and that the two constraints, $\sigma_0$ and $\sigma_1$, are first-class constraints.
The gauge transformations generated by the  the first-class constraint $\sigma_1$ are 
\begin{eqnarray}
\nonumber
\begin{array}{ccc}
\quad A^{'}_0 = A_0 -\partial_0\alpha, \qquad & \qquad\quad A^{'}_1 = A_1 -\partial_1\alpha,\qquad & 
\pi^{'}_1 = \pi_1, \\
 {} & \rho^{'} = \rho, \qquad  & \qquad \pi_\rho^{'} = \pi_\rho, \qquad \\
\varphi^{'} = \varphi + e\alpha, \qquad & \pi^{'}_\varphi =\pi_\varphi, \qquad
& \partial_1\varphi^{'} + e A^{'}_1 =\partial_1\varphi + eA_1,
\end{array}
\end{eqnarray}
$\alpha(t,x)$ being an arbitrary function of spacetime defined modulo $2\pi/e$ and periodic in $x\rightarrow x+2L$.
Besides these small gauge transformations topologically connected to the identity transformation, the system also possesses
an invariance under large gauge transformations leaving all variables invariant except for $A_1$ and $\varphi$
which are shifted in a quantised fashion in such a manner as if the above gauge function $\alpha(t,x)$
took the value $\alpha_\ell(t,x)=x\pi\ell/(eL)$ with $\ell\in\mathbb{Z}$ and $\ell\ne 0$.

\subsection{The physical Hamiltonian formulation}

In order to obtain a physical Hamiltonian formulation of the system,
we need a canonical redefinition of its degrees of freedom. 
Let us then introduce \cite{lau}
\begin{eqnarray}
\nonumber
B=-\frac{1}{e} E,\qquad \pi_B =\partial_1\varphi + e A_1.
\end{eqnarray}
The first-order gauge invariant action then becomes
\begin{eqnarray}
\nonumber
S_{\textrm{phys}} =\int_{\mathbb{R}} dx^0 \int_{-L}^{+L} dx^1
\left[\partial_0 B\pi_B + \partial_0\rho\pi_\rho -{\mathcal{H}}_{\textrm{phys}}\right],
\end{eqnarray}
where 
\begin{equation}
{\mathcal{H}}_{{\textrm{phys}}} = \frac{1}{2}\rho^2\pi_B^2 + \frac{1}{2\rho^2}(\partial_1 B)^2
+\frac{1}{2}e^2B^2
+\frac{1}{2}\pi_\rho^2 +\frac{1}{2}(\partial_1\rho)^2 + V(\rho)
+\partial_0[B\pi_B] -\partial_1\left[ B(\partial_0\varphi +e A_0)\right],
\label{surf}
\end{equation}
any remaining terms then cancelling on account of the gauge constraints.

\subsection{The physical Lagrangian formulation}

The Hamiltonian equations of motion in $\rho$ and $B$ are given by
\begin{eqnarray}
\partial_0\rho &=& \int_{-L}^L dy\left\{ \rho(t,x)\:,\:{\mathcal{H}}_{\textrm{phys}}\right\}
 = \int_{-L}^L dy \left\{\rho(t,x)\:,\:\frac{1}{2}(\pi_\rho)^2(t,y)\right\}
 = \pi_\rho, \label{415} \\
\partial_0 B &=& \int_{-L}^L dy\left\{ B(t,x)\:,\:{\mathcal{H}}_{\textrm{phys}}\right\}
=\int_{-L}^L dy \left\{B(t,x)\:,\:\frac{1}{2}\left(\rho^2(\pi_B)^2\right)(t,y)\right\}
= \rho^2\pi_B. \label{416}
\end{eqnarray}
Considering the equations of motion (\ref{415}) in $\rho$, and (\ref{416}) in $B$,  and
reducing their associated conjugate momenta in the 
expression for ${\mathcal{L}}_{\textrm{phys}} = \pi_\rho\rho +\pi_B B
-{\mathcal{H}}_{phys}$, the Lagrangian formulation of the system in its gauge invariant
physical sector is the following,
\begin{equation}
{\mathcal{L}}_{\textrm{phys}} = \frac{1}{2}\frac{1}{\rho^2}(\partial_\mu B)^2 
-\frac{1}{2} e^2 B^2 +\frac{1}{2}(\partial_\mu\rho)^2 -V(\rho)
 - \partial_0[\frac{1}{\rho^2}B\partial_0 B]
+\partial_1\left[B(\partial_0\varphi +e A_0)\right].
\label{sur}
\end{equation}
Note that even though the analysis was developed within the Hamiltonian formulation,
this expression for the physical Lagrangian density is once again manifestly
Lorentz invariant, up to surface terms.

\subsection{The non-perturbative dynamics in the physical sector}

In this sector, the model coincides with the coupling of a pseudoscalar
field, namely the electric field $E$, to the real scalar Higgs field $\rho$,
which are both neutral and U(1) gauge invariant, since they lie within the physical sector.
All the other and gauge non invariant degrees of freedom have indeed decoupled from the physical Lagrangian.

\subsubsection{The electric field}

Up to its normalisation, the field $B$ is the electric field which acquires a mass proportional
to the gauge coupling constant, $e$. This mass also depends on the vacuum
expectation value of $\rho$. In fact, if $\rho$ is constant, namely such that $\rho(x) =\rho_0$,
the kinetic factor $1/(2\rho^2)(\partial_\mu B)^2$ for $B$ is well defined and, by an appropriate
change of normalisation in relation to the term in $B^2$, corresponds to a pseudoscalar field of mass $|e\rho_0|$.
The reason why the electric field, or $B$ is pseudoscalar is that under the parity transformation
in $1+1$ spacetime dimensions, the electric field changes its sign (while there is no notion of rotation
in a space with a single dimension). 

Let us now assume that $\rho(t,x)$ takes any real field configuration, not necessarily constant nor static.
Since the kinetic energy term for $B$ includes the factor
$ 1/\rho^2$, this implies that the field $B$ acquires a 
dynamical mass of which the value depends on the configuration $\rho(t,x)$ for
the physical scalar field. Incidentally, it thus also follows that the system
classifies as an effective representation for
superconductivity, and indeed this model corresponds to the celebrated
Laudau--Ginzburg model in the stationary case when extending the number of space dimensions,
or the Higgs model in the case of the $U(1)$ gauge symmetry in higher dimensions where 
the gauge boson then acquires a dynamical mass.

\subsubsection{The radial field}

The radial field $\rho$ arises as a real field in the theory, while
its mass value, as does that of the electric field, depends on the choice of scalar potentiel, $V(\rho)$.
A choice of interest is the Higgs potential of the form
\begin{eqnarray}\nonumber
V(\rho) =\frac{1}{8} M^2(\rho^2 -\rho_0^2)^2,
\end{eqnarray}
where $M^2$, with $M>0$, is a factor that possesses the physical 
dimension of a squared mass, and $\rho_0>0$ is the vacuum expectation value of the Higgs sector.
Given this choice, the mass of the Higgs field $\rho$ is given by $ |M\rho_0|$.

\subsection{Noether symmetries}

According to the Noether theorem \cite{Gov1}, applied to the present system
in the case of spacetime translations,
\begin{eqnarray}
\nonumber
x'^\mu =x^\mu +a^\mu,\qquad
\rho'(x') =\rho(x),\qquad
B'(x') = B(x),
\end{eqnarray}
it follows that the current density is given by,
\begin{eqnarray}
\nonumber
\gamma^{\nu\mu} =\partial^\mu\rho\partial^\nu\rho
 +\frac{1}{\rho^2}\partial^\mu B\partial^\nu B -\eta^{\nu\mu}
{\mathcal{L}}_{phys},
\end{eqnarray}
an expression thus defining the energy-momentum tensor
of the system in its physical sector.
The associated conserved charges are the total energy, $E_{\rm phys}$, and
momentum, $P_{\rm phys}$, of the system. One finds
\begin{eqnarray}
\nonumber
E_{\rm phys} = \int_{-L}^L dx\left[\frac{1}{2\rho^2}(\partial_0 B)^2
+\frac{1}{2\rho^2}(\partial_1 B)^2 +\frac{1}{2}e^2 B^2
+\frac{1}{2}(\partial_0\rho)^2 +\frac{1}{2}(\partial_1\rho)^2 + V(\rho)\right],
\label{444}
\end{eqnarray}
and
\begin{eqnarray}
\label{445}
P_{\rm phys} = -\int_{-L}^L dx \left[ \partial_0\rho\partial_1\rho  + \frac{1}{\rho^2}
\partial_0 B\partial_1 B\right].
\end{eqnarray}

Furthermore, we have the Noether current associated to the local $U(1)$ gauge symmetry,
\begin{eqnarray}
\nonumber
j_\mu &=& -\frac{1}{2}i\frac{1}{|\phi|^2}\left[\phi^{*}D_\mu\phi
-(D\mu\phi)^{*}\phi\right]\\\nonumber
&=&-\frac{1}{2}i\frac{1}{\rho^2}\left[\rho\left(\partial_\mu\rho
+i\rho(\partial_\mu\varphi +eA_\mu)\right)
-\rho\left(\partial_\mu\rho-i\rho(\partial_\mu\varphi +eA_\mu)\right)\right]\\\nonumber
&=&\partial_\mu\varphi +e A_\mu, 
\end{eqnarray}
as well as the identity
\begin{eqnarray}
\nonumber
\epsilon^{\mu\nu}\partial_\mu[B j_\nu] &=& \partial_0\left[B(\partial_1\varphi +eA_1\right]
-\partial_1\left[B(\partial_0\varphi +eA_0)\right]\\\nonumber
&=&\partial_0[B\pi_B]-\partial_1[B(\partial_0\varphi +eA_0)].
\end{eqnarray}

\section{Concluding Remarks}
\label{Sec4}

We have thus reached the following fundamental conclusion: at the classical level,
the two dimensional scalar electrodynamics model, in the symmetry breaking phase,
is physically equivalent to a coupled system of a neutral pseudo-scalar field
and a neutral scalar field, both massive.

The Euler-Lagrange equations of motion in $\rho$ and $B$ are,
respectively,
\begin{displaymath}
\partial_\mu^2\rho + \frac{1}{\rho^3}(\partial_\mu B)^2 +
\frac{\partial V}{\partial\rho} =0,\qquad
\partial_\mu[\frac{1}{\rho^2}\partial^\mu B] + e^2 B =0.
\end{displaymath}
These equations constitute a system of coupled non-linear equations.
We are interested in the configurations which are solutions to these 
equations \cite{Brihaye}. It might be possible to have some analytical solutions.

\section*{Acknowledgements}

This work, part of Laure Gouba's Ph.D. thesis, was revisited
during her stay at the African Institute for Mathematical Sciences (AIMS)
as a Postdoctoral Fellow and Teaching Assistant. L.G. would like to
thank Profs. Neil Turok, founder of AIMS, and Fritz Hahne, Director
of AIMS, as well as the AIMS family for their support and hospitality.

J.G. is grateful to Profs. Hendrik Geyer, Bernard Lategan and Frederik Scholtz for the support and the hospitality
of the Stellenbosch Institute for Advanced Study (STIAS) with the grant of a Special STIAS Fellowship
which made a stay at STIAS and NITheP possible in April-May 2008. He acknowledges the Abdus Salam International
Centre for Theoretical Physics (ICTP, Trieste, Italy) Visiting Scholar Programme in support of
a Visiting Professorship at the ICMPA-UNESCO (Republic of Benin).
J.G.'s work is also supported by the Institut Interuniversitaire des Sciences Nucl\'eaires, and by
the Belgian Federal Office for Scientific, Technical and Cultural Affairs through
the Interuniversity Attraction Poles (IAP) P6/11.


\begin{thebibliography}{99}

\bibitem{lau}
L. Gouba, {\it Th\'eories de jauge ab\'eliennes scalaire 
et spinorielle \`a $1+1$ dimensions: Une \'etude non perturbative},
Ph.D. Thesis (University of Abomey--Calavi, Benin, 2005), unpublished.

\bibitem{Gov1}
J. Govaerts, {\it Hamiltonian Quantisation and Constrained Dynamics}
(Leuven University Press, Leuven, 1991).

\bibitem{Brihaye}
Y. Brihaye, S. Giller, P. Kosinski and J. Kunz, {\sl Phys. Lett. B} {\bf 293}, 383 (1992).

\end{thebibliography}
\end{document}